\documentclass{emulateapj}

\usepackage{hyperref}
\hypersetup{colorlinks=true,citecolor=blue}
\usepackage{natbib}
\bibpunct{(}{)}{;}{a}{}{,} 

\shorttitle{The X-ray eclipse geometry of CAL~87}
\shortauthors{T. Ribeiro}

\defcitealias{2007A&A...472L..21O}{OS07}

\begin{document}

\title{The X-ray eclipse geometry of the super-soft X-ray source CAL~87}

\author{T. Ribeiro and R. Lopes de Oliveira}
\affil{Universidade Federal de Sergipe, Departamento de F\'isica, Av. Marechal Rondon s/n, 49100-000 S\~ao Crist\'ov\~ao, SE, Brazil}
\email{T. Ribeiro: tribeiro@ufs.br, R. Lopes de Oliveira: rlopes@ufs.br}

\and

\author{Borges, B. W.}
\affil{Universidade Federal de Santa Catarina, Campus Ararangu\'a, 88905-120, Ararangu\'a, SC, Brazil}
\email{bernardo@astro.ufsc.br}

\begin{abstract}

We explore XMM-{\it Newton} observations of the eclipsing super-soft X-ray source CAL~87 in order to map the accretion structures of the system. { Indirect imaging techniques were applied in X-ray light curves to provide eclipse maps}. The surface brightness distribution exhibits an extended and symmetric emission, { and from the hardest X-rays is revealed a feature that is likely due to a bright spot}. A rate of $\dot{P} = (+6\pm2) \times 10^{-10}$ for changes in the orbital period of the system was derived from the eclipses. There is no significant variation of the emission lines even during eclipses, arguing that the lines are formed in an extended region. The continuum emission dominates the decrease in flux which is observed during eclipses. The O\,{\small VIII} Ly$\alpha$ line reveals a broadening velocity { which is} estimated in 365$^{+65}_{-69}$ km\,s$^{-1}$ (at 1$\sigma$) and marginal evidence for asymmetry in its profile, and sometimes shows evidence of double-peaked emission. Together, the results support that the wind-driven mass transfer scenario is running in CAL~87.

\end{abstract}


\section{Introduction}
\label{sct:intro}

Compact binaries super-soft X-ray sources (CBSS) are systems in which an evolved compact component, e.g. white-dwarf (WD), accretes matter from its companion up to Eddington limit (see \citealt{1997ARA&A..35...69K} for a review).  The high mass-transfer ratio would lead to stable nuclear burning in the surface of the WD, explaining the super-soft X-ray emission of these systems \citep{1992A&A...262...97V}. In order to { explain} the high mass-accretion rates required to describe the X-ray emission in CBSS, \cite{1992A&A...262...97V} proposed that accretion from inverted mass-ratio systems would take place, in the sense that the more massive component is the donor star.  This phenomenon would lead to a decrease in the orbital separation of the system, shrinking the Roche lobe of the star, thus forcing more matter to flow from the donor { to its companion}. { This} star then become thermally unstable and mass-transfer occurs on Kelvin-Helmholtz time-scales. 

On a number of situations though, this dynamical instability mass-transfer (DIMT) scenario has been brought into question. For instance, surface nuclear burning, required to explain CBSS, assumes massive WD which means that systems with orbital period less than about 6~hr cannot harbor more massive companions \citep{1992A&A...262...97V}.

Testing the DIMT framework would simply require measuring the mass-ratio of CBSS systems. Nevertheless, as the disk and boundary layer resulting from the high mass-accretion rates dominate the emission -- practically at all wavelengths -- it is almost impossible to detect their donor component. Therefore, it is challenging to determine their orbital parameters, even for eclipsing systems. 

A second outcome of DIMT is that the systems must display a considerable decrease in orbital period, as a result of the decrease in the orbital separation \citep{1997ARA&A..35...69K}. In this case, monitoring eclipse timings in highly inclined systems enable the determination of orbital period changes. 

There are now at least two CBSS systems which are clearly not explained by the DIMT model: \object{SMC~13} and, the target of this work, \object{CAL~87}. SMC~13 (or \object{CBSS 1E0035.4-7230}) is a short-period ($P_{orb} = 4.1 \rm h$) system in the Small Magellanic Cloud with a mass-ratio lower than unity \citep{2000A&A...355.1041G}. On the other hand, CAL~87 is an eclipsing CBSS in the Large Magellanic Cloud with a longer orbital period ($P_{orb} = 10.6 \rm h$), above the $6\rm hr$ limit expected for the DIMT model. Nevertheless, using optical photometric observations spanning about $\sim 17 \rm yr$, \citet[hereafter \citetalias{2007A&A...472L..21O}]{2007A&A...472L..21O} inferred an orbital period increase of $\dot{P} = (+1.7\pm0.3) \times 10^{-10}$, thus challenging the DIMT explanation for CAL~87.

Alternatively, \cite{1998A&A...338..957V} proposed a wind-driven mass transfer (WDMT) model to explain the high mass-transfer rates expected in CBSS. In this case, strong stellar winds could be induced if the donor star is heavily irradiated by soft X-rays from its companion. At the same time, since mass would be transferred from the less massive to the more massive component, this scenario is consistent with the reported orbital period increase in CBSS-like objects (\citealt{1998PASP..110..276S} and \citetalias{2007A&A...472L..21O}). Nevertheless, direct observations of the irradiated companion is again problematic since the emission from the system is dominated by the disk and other accretion-related structures. 

In this work we analyze XMM-{\it Newton} observations of CAL~87 with aid of eclipse mapping techniques \citep{1985MNRAS.213..129H} in order to map and characterize the X-ray accretion structures of the system. The investigation is complemented by qualitative analysis of X-ray light curves and spectra.

\section{Observations}
\label{sct:obs}

CAL~87 was the main target of a 78 ks { campaign} carried out by the XMM-{\it Newton} observatory on 2008 April 18 (PI: Ken Ebisawa; ObsID 0153250101). We report here results from such a { campaign, exploring the EPIC and RGS data}. All EPIC (pn, MOS1, and MOS2) cameras were operated in the {\it full frame} mode, implying in a time resolution of 2.6 s for both MOS cameras and 73.4 ms for the pn camera. A thin optical blocking filter was used for each EPIC camera. The RGS1 and RGS2 data were accumulated in the ``spectroscopy'' mode. There is no significant photon pile-up in the EPIC and RGS data. High background contamination has only slightly affected the end of the observations; such an interval was excluded only for the spectroscopic investigations, resulting in a good time interval of 48.6 ks for the pn, 61.2 ks for the MOS1/2, and 76.2 ks for the RGS1/2 cameras. The data were reprocessed ({\it epproc}, {\it emproc}, and {\it rgsproc} tasks) and reduced with the Science Analysis System (SAS) software v13.0 using standard procedures with the calibration files available in 2013 December 11. The background dominates the emission for E $\ga 1.4 \rm keV$. For the light curves, always background subtracted ({\it epiclccorr} task) and with times corrected to the barycenter of the solar system ({\it barycen} task), we assume three energy bands: 0.3-0.5 keV and 0.5-1.4 keV, hereafter the ``soft'' and ``hard'' X-ray bands, respectively, and 0.7-1.4 keV band. We use Xspec\footnote{http://heasarc.nasa.gov/docs/xanadu/xspec/index.html} v12.8.0 to carry spectral fits and construct fake spectra, applying the following standard models: {\it bbody} for black-body emission, {\it phabs} to account for the photoelectric absorption, and {\it edge} for absorption edges.

\begin{figure}
\centerline{
\includegraphics[scale=0.5,angle=270]{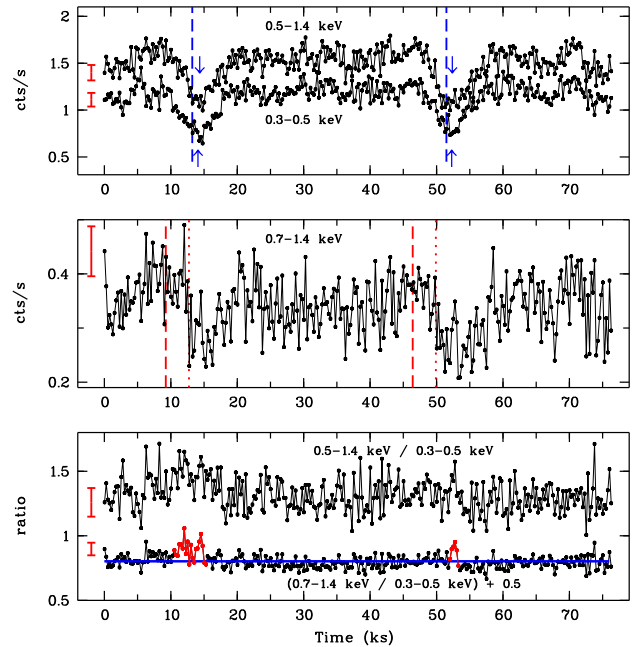}
}
\caption{X-ray light curves (top and middle) and hardness ratio (bottom) of CAL~87. The upper limits for the error bars are showed in the left of each curve. In the top panel, the {\bf vertical} dashed lines indicates the time of minimum light, according to Eq.\ref{eq:efem}, and arrows indicate the corresponding measurements (see text for discussion). In the middle panel, the {\bf vertical} lines indicates the start of the eclipses in the ``soft" (dashed) and the 0.7-1.4 keV (dotted) X-ray bands. {\bf The horizontal line in the bottom panel represents the mean value for the hardness ratio 0.7-1.4 keV / 0.3-0.5 keV, and the red parts emphasize the suspect increase in the hardness during eclipse.} \label{fig:clfold}}
\end{figure}

The data were phase folded according to the linear ephemeris of \cite{1997MNRAS.287..699A};
\begin{equation}
\label{eq:efem}
T_0({\rm E}) = {\rm HJD}\, 2\,45\,0111.5144(3)+0.4426\,7714\,(6) {\rm E}, 
\end{equation}
{ where $\rm E$ is the cycle and $T_0$ the time of the eclipse.} This ephemeris considers inferior conjunction -- where the secondary is in front of the primary -- as the phase of optical (V-band) minimum light. Also it is important to note that Eq.~\ref{eq:efem} considers cycle 0 as the one defined by \cite{1997MNRAS.287..699A} and not that of \cite{1989MNRAS.241P..37C}, the first published ephemeris. In order to fix the cycle counting we added 5884 to E (see discussion furthermore). In Fig.~\ref{fig:clfold} we show light curves of CAL~87 from three energy bands (0.3-0.5 keV, 0.5-1.4 keV, and 0.7-1.4 keV), and in Fig.~\ref{fig:emm} the phase folded light curve centered in the eclipses.

\section{Data Analysis}
\label{sct:data}

{ The XMM-{\it Newton} observations of CAL\,87 covered two eclipses of the system.} Overall, the profile of the X-ray light curve of CAL~87 is strikingly similar to that observed in the optical (e.g., \citealt{1989MNRAS.241P..37C} or \citealt{1998ApJ...502..408H}, and references therein, for more recent data). Nevertheless, the phase of minimum light observed in the X-ray is slightly off-set to later phases, as can be seen in Fig.~\ref{fig:clfold} ({ vertical lines in the top panel}). This phase shift of the X-ray minimum was previously reported by \cite{2004RMxAC..20..210O} using { the same X-ray dataset considered in this work}. On the other hand, \cite{1993PASP..105..863S}, using a different X-ray dataset (ROSAT/PSPC), reported that the time of minimum brightness in X-rays coincided with those in the optical.

We measured the time of minimum light in X-rays using the same procedure applied by \cite{1997MNRAS.287..699A}, fitting a parabola to the most symmetrical central part of the eclipse. Since our data cover two eclipses entirely, the fit was repeated independently for each one of them. The resulting times of minimum light, together with those predicted by  Eq.~\ref{eq:efem}, and the corresponding phase displacements, are shown in Table~\ref{tab:efem}.  The uncertainties were estimated by means of a Monte-Carlo simulation. { The arrows and vertical lines in the top panel of Fig.~\ref{fig:clfold} indicates the measured and the computed values, respectively, for the minimum light.}


\begin{table}
\begin{center}
\caption{\label{tab:efem} Times of x-ray minimum light}
\begin{tabular}{ccccc}
\tableline\tableline
Band & Cycle$^1$ & $T_0$ calculated & $T_0$ measured & $\Delta \phi$ \\
\tableline
Soft & 11841 &  2452748.5421(3)  & 2452748.552(2)   & $+0.024(5)$  \\   
Soft & 11842 &  2452748.9848(3) & 2452748.994(2) & $+0.020(4)$ \\   
Hard & 11841 &  2452748.5421(3) & 2452748.555(2)   & $+0.030(5)$  \\   
Hard & 11842 &  2452748.9848(3) & 2452748.995(2) & $+0.023(3)$ \\   
\tableline
\tableline
\end{tabular}
\end{center}
$^1$ Considering cycle 0 that of \cite{1989MNRAS.241P..37C}. In order to calculate the timings of eclipsing with Eq.~\ref{eq:efem}, one needs to subtract 5884 from the cycle counting.
The ``soft" and ``hard" bands are integrated between 0.3 - 0.5 keV and 0.5 - 1.4 keV, respectively.
\end{table}

The resulting phase displacements { with respect to the linear ephemeris} measured for the two observed X-ray eclipses are significant and mutually consistent ($\Delta \phi \sim 0.02$, $\Delta T \sim 13 \, {\rm min}$). Since the eclipse of the compact central source itself (possibly a WD) is not visible in the light curve, it is difficult to pinpoint the origin of { such a displacement}. The explanation should be, more likely, either a true anomaly in the orbital period of the system, causing the phase of minimum to digress, or a non-uniform surface brightness distribution of the eclipsed source.  

Orbital period changes are indeed very common on eclipsing systems and occur on timescales of years to decades \citep[and references therein]{2008A&A...480..481B}. In fact, \citetalias{2007A&A...472L..21O} already reported orbital period variations in CAL~87, from optical data. 
{  
One may argue that combining timings in the optical with those in the X-rays is problematic, given that the geometry of the eclipses may be quite different. Nevertheless, as shown by \cite{1993PASP..105..863S} both timings agree well for contemporaneous observations, suggesting that the combination of optical and X-ray timings may be possible. With that in mind, we attempted to reproduce \citetalias{2007A&A...472L..21O} results including the timings measured with our X-ray data.
} 

Initially we attempted to estimate the magnitude of the orbital period changes by measuring the time between eclipses. This yields in: 
\begin{equation}
\label{eq:porbdet1}
{\rm T_0(11841)} - {\rm T_0(11840)} = P_{orb} = 0.441 \pm 0.004 \rm d.
\end{equation} 
Values obtained with other techniques are also consistent with this result. A phase dispersion minimization \citep{1978ApJ...224..953S} resulted in $P = 0.435 \pm 0.009 \, {\rm d}$, and a Lomb-Scargle Periodogram gives only a marginally significant peak at $P \sim 0.437 \, \rm d$. Unfortunately, with the available data, it is not possible to independently determine the orbital period of the system with sufficient precision to measure orbital period changes, as performed by \citetalias{2007A&A...472L..21O}. 

Even though we are unable to directly measure orbital period changes with this procedure, we can at least estimate the required $\dot{P}$ to produce the observed phase shift. For that we consider a steady change in $P$ ($\dot{P}$ constant), which gives a modified ephemeris of the form:
\begin{equation}
\label{eq:modefem}
T'_0({\rm E}) = {\rm HJD_0}+P \cdot {\rm E} = {\rm HJD_0}+P_0 \cdot {\rm E} + \dot{P} \cdot \Delta T \cdot {\rm E}. 
\end{equation}
Note that $T_0({\rm E}) = {\rm HJD_0}+P_0 \cdot {\rm E}$ is the usual linear ephemeris (Eq.~\ref{eq:efem}) and $P = P_0 + \dot{P} \cdot \Delta T$ account for the varying orbital period of the system. Here $\Delta T = T'_0({\rm E})-{\rm HJD_0}$, so the ephemeris in this form is a transcendental equation and must be solved iteratively. Since we are interested in $\dot{P}$, we re-write Eq.~\ref{eq:modefem} in the form:
\begin{equation}
\label{eq:modefem2}
T'_0({\rm E}) - T_0({\rm E}) = \dot{P} \cdot {\rm E} \cdot (T'_0({\rm E})-{\rm HJD_0}), 
\end{equation}
or
\begin{equation}
\label{eq:modefem3}
\dot{P}  = \frac{1}{{\rm E} } \cdot \frac{\Delta T'_0}{\Delta T} , 
\end{equation}
where $\Delta T'_0 = T'_0({\rm E}) - T_0({\rm E})$ is the difference in time between the observed and the calculated minimum light, and ${\Delta T} = T'_0({\rm E})-{\rm HJD_0}$ is the difference between the time of the first ephemeris and the current ephemeris. Since $\Delta T'_0$ is measured, we solve this iteratively, starting by considering $T'_0({\rm E}) = T_0({\rm E})$ to obtain ${\Delta T}$, calculate $\dot{P}$, then, calculate a modified $T'_0$, which yields a new ${\Delta T}$ and so on. The procedure is repeated until the calculated $\dot{P}$ does not change considerably. Using this procedure we obtain $\dot{P} = (+6\pm2) \times 10^{-10}$.

\subsection{Eclipse mapping}
\label{ssec:eclmap}

We investigate the geometry of the eclipse of CAL~87 by means of the eclipse mapping method (hereafter EMM, \citealt{1985MNRAS.213..129H}). This technique works by solving an indirect imaging inversion problem, to provide a surface brightness map that best fit the eclipse profile. It assumes that all changes in flux are caused solely by the eclipse of a flat 2-dimensional surface plus an uneclipsed component which does not vary in phase. Since performing an inversion from a 1-dimensional dataset to a 2-dimensional map yields a number of non-unique solutions, an entropy regularization scheme is employed in order to select the most uniform and featureless surface brightness distribution \citep{1993A&A...277..331B}.

{ 
The EMM technique was applied to the soft and hard X-ray band light curves, separately. The corresponding input light curves were previously phase-folded assuming the new ephemeris shown in Section~\ref{sct:data}, combining thus data covering two eclipses, and finally considering only the phase interval [-0.2,0.2] as input for the EMM.} In Fig.~\ref{fig:clfold} it is possible to note that there are small out-of-eclipse variations in flux. More importantly, we note that the flux level before and after the eclipses are slightly different. Since the EMM fits only the eclipses, this contribution must be subtracted. We proceeded by fitting a 3rd order polynomial function to the out-of-eclipse data (for the phase intervals [-0.2:-0.1] and [0.1:0.2]), dividing the light curve by the resulting polynomial, and scaling the result to the value at phase zero of the function. In this case, we found that a 3rd order polynomial function is more suitable than a spline-fit (e.g., \citealt{2007AJ....134..867B}), avoiding high-degree interpolation oscillations.

In addition to the phase-folded light curve, the EMM requires knowledge of the system parameters namely, the inclination ($i$) and mass-ratio\footnote{In this work we consider that $M_2$ is the mass of the donor.} ($q = M_2 / M_1$) of the system. Even though CAL~87 is a well known eclipsing system (which usually enables precise determinations of $i$ and $q$), there is no consensus on the determination of these parameters. \cite{1992A&A...262...97V} and \citet{1997A&A...318...73S} suggested that the system is well described by a more massive ($M_2 \sim 1.5 M_\odot$, evolved F - G star) secondary and a less massive ($M_1 \sim 0.75 M_\odot$, possibly WD) primary.  Being this the case, $q \sim 2.0$. 

Furthermore, \cite{1996BAAS...28..923W} show from HST observations that the F-G colors observed during primary eclipses -- which were used to infer the spectral type of the secondary -- are probably of a nearby ($0.9"$) field G star. Later \cite{2004ApJ...612L..53S} used a model for surface Hydrogen-burning to  show that the primary is more likely a massive ($M_1 = 1.75 M_\odot$) WD. This result was later confirmed by radial velocity measurements \citep{1998ApJ...502..408H}, which are consistent with a more massive primary ($M_1 \sim 1.4 M_\odot$) and a less massive secondary ($M_2 \sim 0.34 M_\odot$), hence $q \sim 0.25$. 

Although the eclipses allow to constrain the value of inclination of the system to a well defined range ($i \gtrsim 70^\circ$), there is still no reliable determination of it. In general, for CAL~87, it is adopted the inclinations obtained by either \citet{1997A&A...318...73S} or \citet{1997A&A...321..245M}, $i = 77^\circ$ and $i = 78^\circ$ respectively. Even tough they inferred drastically different values for mass ratio ($q=2.0$ and $q=0.25$, respectively), their values for the inclination of the system are virtually the same. 
We then proceed to perform the EMM adopting the more recent (and, allegedly, more precise) set of system parameters, namely, $q=0.25$ and $i = 79^\circ$ \citepalias{2007A&A...472L..21O}. 

Adopting these values, the resulting surface brightness reconstruction displayed an extended and asymmetric feature between both stellar components of the system. However, such features are usually observed on EMM reconstruction when the inclination of the system is underestimated. 
In order to overcome this issue we investigated the impact of the choice of $i$ in the surface brightness reconstruction. For that we repeated the EMM with $i$ varying from $75^\circ$ (the lower inclination that result in eclipses) to $85^\circ$, in steps of $0.25^\circ$, inspecting the final value of the entropy. Higher values of entropy results in featureless and more reliable reconstructions. We observe an asymptotic behavior for the entropy of the reconstructions with increasing inclination, stabilizing for $i > 82^\circ$ with the disappearance of the asymmetric features previously observed.

Even though we were able to use the entropy of the reconstruction to find a more suitable value for the inclination, the situation is more complex when it comes to constraining the mass ratio. Instead of introducing spurious artifacts on the image reconstruction, the mass ratio affect the relative sizes of the Roche lobe of the components, reflecting in the description of the depth and width of the eclipse. As discussed by \citet{1996MNRAS.282...99B}, there is a correlation between this two quantities limiting the ability of the method to constrain the relative sizes of the Roche lobe, e.g., the mass ratio. 

In order to access the impact of the choice of the mass ratio in the reconstruction, we performed the EMM with different mass ratios ranging from $0.1$ to $2.5$, in steps of $0.05$. The reconstructions ends with similarly high entropy values, making it impossible to select the one which resulted in the most featureless surface brightness. As expected, we observe a steep increase in the uneclipsed component for larger values of mass ratio. Clearly, there is a counter-balance between an increase in the relative size of the Roche lobes -- larger values of $q$ results in relatively larger Roche lobe of the donor -- and an increase in the uneclipsed component. We then decided to adopt the mass ratio value proposed by \citet[$q=0.25$]{1998ApJ...502..408H}. In any case, we point it out that for mass ratios of the same order of magnitude ($0.15 - 0.3$) we do not observe any substancial change in the surface brightness reconstruction.

\begin{figure*}[htbp]
\centerline{
\includegraphics[scale=0.75]{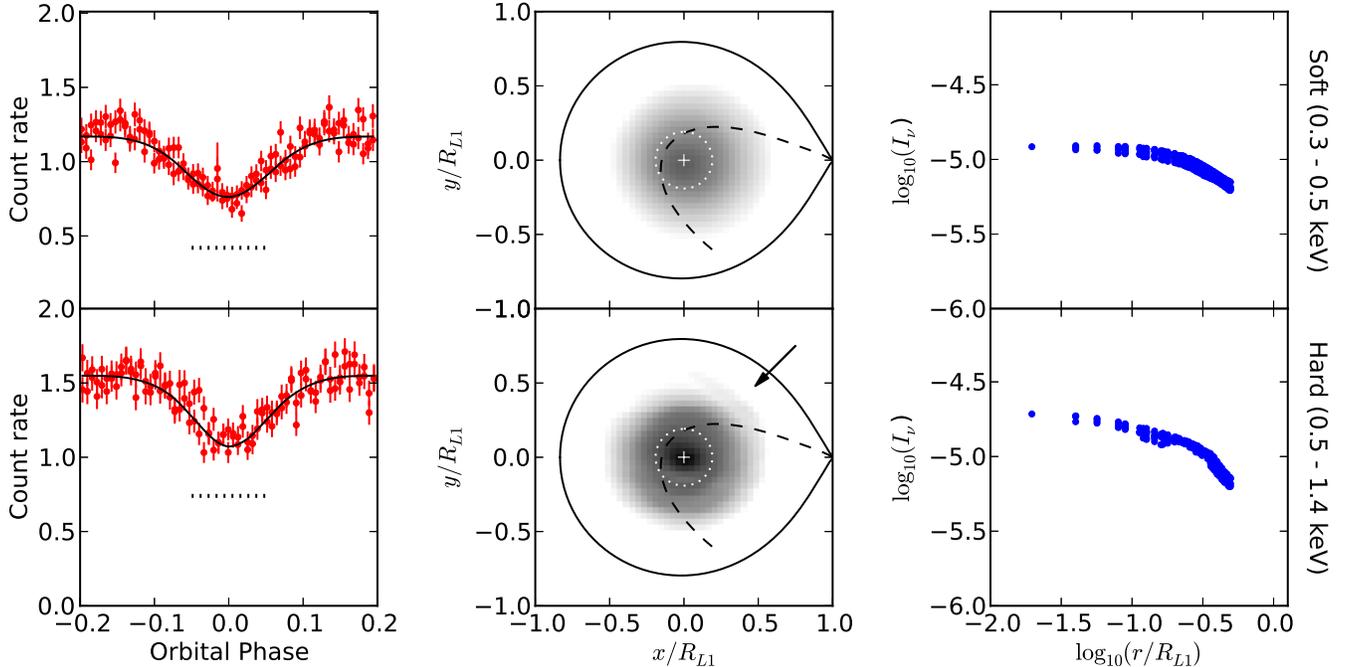}
}
\caption{Soft (top) and hard (bottom) X-ray bands eclipse mapping results for CAL~87. The left panels show the phase-folded light curves of the system (dots with error bars) together with the resulting model (solid lines) { and the uneclipsed component (horizontal dotted line)}. The gray-scaled maps at the center represents the resulting surface brightness distribution in a logarithm gray-scale (same scale for both X-ray bands). The Roche lobe of the accreting component is represented in solid lines, the ballistic trajectory is shown is dashed lines and the circularization radius for a mass-ratio of $q=0.25$ in white dotted lines (see text for discussion). A cross marks the center of the disk and an arrow in the bottom plot depicts the asymmetric feature we identify in the ``hard" X-ray surface brightness image. The right hand panels show the radial brightness distribution for each map. \label{fig:emm}}
\end{figure*}

Finally we adopted $i=82^\circ$ and $q=0.25$ for the system. The results of the reconstruction for the ``soft'' and ``hard'' X-ray light curves are shown at the top and bottom of Fig.~\ref{fig:emm}, respectively. The left hand panel of the figure shows an inset of the dataset around the eclipses (data points with error bars) together with the resulting model light curve (solid line). { The uneclipsed component level is shown as an horizontal dotted line.} The surface brightness distribution is displayed in the middle panel as a 2-dimensional gray-scaled map in logarithmic scale, overploted with the position of the Roche lobe of the accreting component (solid line), the ballistic trajectory (dashed line), and the circularization radius (white dotted line). The right hand panel shows the radial brightness distribution.

Two results emerge from our analysis. First, the X-rays come from an extended and symmetric surface brightness distribution. For the soft X-ray band the distribution is more flat at the center, rapidly fading for radius larger than $0.5 \rm R_{L1}$. On the other hand, the hard X-ray band has a brighter feature at the center of the map, inside the circularization radius, and steadily fades for outer regions. Second, there is a feature at the edge of the disk for the hardest X-ray map (indicated by an arrow in Figure~\ref{fig:emm}), close to the intersection with the ballistic stream trajectory.

{ 

\citet{2004ApJ...615L.129K} showed that systems with high mass-accretion rates, such as in CAL~87, may have thick disks. This could result in self obscuration and, in the most extreme cases, the only visible feature of the accretion disk would be a thick wall. The departure of a thin disk geometry - one of the main assumption of the EMM - will most certainly affect the surface brightness reconstruction. If only the outer edges of the disk are  obscured, the result should be a flatter radial brightness distribution. This is basically a geometrical issue, as we see the surface of the disk more inclined and the edge of the obscuring wall more face-on. The wall then appears brighter than the regions it obscures.  In this case, we can still map the unobscured regions of the disk. 

If, on the other hand, the entire disk is self-obscured by its outer edge we loose all ability to map its inner features. If the wall is featureless the resulting surface and radial brightness distribution should be flat. If there is a hot-spot (or any other bright feature), there will be a strong orbital modulation caused by the changing viewing angle \citep{2007A&A...474..213R}.

The surface and radial brightness distribution for the soft X-ray band does indeed resemble the expected result for a partially self-obscured accretion disk; featureless surface brightness and flat radial profile. The case is completely opposed for the hard X-ray band, though. Here we are able to map a feature at $\sim 0.5 \rm R_{L1}$ and the radial distribution is much more inclined, suggesting that a thin disk distribution is more likely the case. 

It is important to note that the geometry of the disk will be strongly wavelength dependent. Given the high mass-accretion rate it is likely that the disk will have an inflated geometry. But this will be more evident in the optical and in the near-infrared for which the outer and cooler parts of the disk will be more prominent. In the X-rays we are probably seeing a flat disk, owing to its inner and hotter parts, and we do not discard some self obscuration from those outer regions.

}

\subsection{X-ray spectroscopy}
\label{ssec:xrspct}

Current models fail to face the complex signatures seen in high resolution X-ray spectra of SSS \citep[see][for a detailed discussion]{Ness2013}. This is clearly the case for CAL~87, even for the medium resolution spectroscopy explored in this work as discussed below. Despite of this, the rich XMM-{\it Newton} spectra of CAL~87 in high (RGS) and moderate (EPIC) resolution allowed us to infer -- mainly from a qualitative approach -- a number of issues which are relevant for the context of this work.

\begin{figure}[htbp]
\centerline{
\includegraphics[scale=0.35,angle=270]{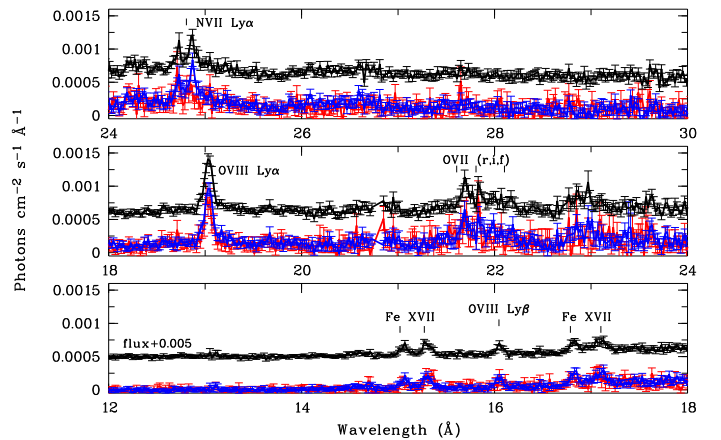}
}
\caption{High resolution X-ray spectra (RGS) of CAL~87 from the whole observation (in black), considering only events collected during eclipses (in red) and from events out-of-eclipses (blue). \label{fig:spec}}
\end{figure}

\begin{figure}[htbp]
\centerline{
\includegraphics[scale=0.4]{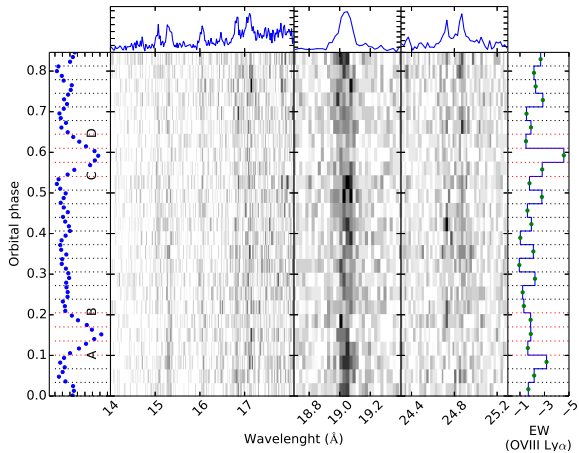}
}
\caption{High resolution X-ray spectra (RGS) around the main lines identified in Fig.~\ref{fig:spec}. They are showed as a function of the time with respect to the light curve in the left, where the eclipses are marked by lines flagged by A-B and C-D. In the right is shown the equivalent width (in \AA) estimated for the O\,{\small VIII} Ly$\alpha$ line. At the top, it is showed the spectra integrated over time for the whole observation, in arbitrary units, for each spectroscopic window. \label{fig:trailedspec}}
\end{figure}

\begin{figure}[htbp]
\centerline{
\includegraphics[scale=0.35,angle=270]{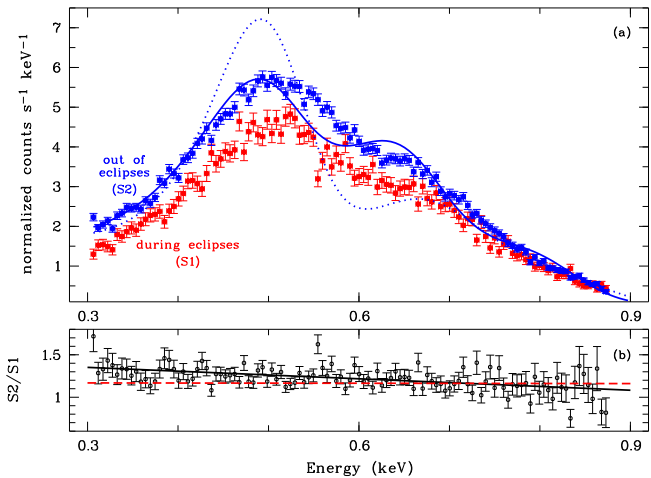}
}
\caption{(a): Medium resolution X-ray spectra (pn) from events collected during times of eclipse (S1; in red) and out-of-eclipses (S2; in blue). { The lines describe the (unacceptable) spectral fit for S2 from simple models as explained in Section 3.2.} (b): The dots represent the S2/S1, with linear regression represented by the solid line. The dashed line represents the linear regression of S2/S1 from fake pn spectra assuming only changes in the flux (see Section \ref{ssec:xrspct}). \label{fig:pn}}
\end{figure}

The high resolution spectrum from the whole observation is shown in Figure~\ref{fig:spec} (in black; added to a constant for clarity). The figure shows also two other spectra: one extracted during times of eclipse (in red; from the A-B and C-D intervals in Fig.~\ref{fig:trailedspec}), and another from events out of eclipse (in blue) -- neither was added to constant values. All three spectra were obtained with the {\it rgsfluxer}/SAS task from the combination of the first and second orders of the RGS1 and RGS2 cameras. The corresponding spectra from the pn/EPIC data are presented in Fig.~\ref{fig:pn}a, named as S1 for times of eclipse and S2 for the rest of the time.

The model applied by \citet{Asai98} and \citet{Ebisawa01} in ASCA observations does not fit the XMM-{\it Newton} data. This is illustrated by the dotted line in Fig.~\ref{fig:pn}a for the pn spectra associated to times out of eclipse. It is based in a composition of black-body ($k \rm T\,\sim\,0.53\, \rm keV$) plus absorption edges components (at 0.739 and 0.871 keV), affected by a photoelectric absorption equivalent to a Hydrogen column (N$_{\rm H}$) of 7.7$\times$10$^{21}$ cm$^{-2}$. The model is even unacceptable when all parameters are free during the fits ($\chi^2_{red}$$\,\sim$\,6.4; solid line in Fig.~\ref{fig:pn}a). For such a case, the photoelectric absorption and temperature of the black-body converge to N$_{\rm H}$\,$\sim$\,3.4$\times$10$^{21}$ cm$^{-2}$ and $\sim$\,103\,keV, respectively, with edges at 0.68\,keV ($\tau$\,$\sim$\,1.5) and 0.83\,keV ($\tau$\,$\sim$\,10). \citet{Ebisawa10} added 12 emission lines in the model of \citet{Ebisawa01}, improving the fit to $\chi^2_{red}$=1.8. The temperature for the black-body emission was $\sim$\,88\,eV. The authors argued that the emission lines arise from an optically thin component. We were not able to improve the fit adding optically thin models as {\it apec} and {\it mekal}. As said previously, our conclusions from the X-ray spectroscopy are based on a qualitative analysis of the data. 

The X-ray spectrum of CAL~87 is marked by several emission lines, absorption edges, and a continuum (Fig.~\ref{fig:spec}). The  prominent lines are those of the N\,{\small VII} Ly$\alpha$, O\,{\small VIII} Ly$\alpha$, and O\,{\small VIII} Ly$\beta$ transitions, and others of Fe\,{\small XVII}.  There is evidence of double-peaked profile that ``comes and goes'' in a non-coherent way for the strong and always present O\,{\small VIII} Ly$\alpha$ line. Figure~\ref{fig:trailedspec} shows the time evolution of a few lines, but only the O\,{\small VIII} Ly$\alpha$ line has signal-to-noise which is good enough for this analysis. The “trails” which are visible in the figure follow the time evolution, with the widths representing the time intervals adopted to accumulate the photons for each case. The O\,{\small VIII} Ly$\alpha$ line reveals two other important features: evidence of non-symmetrical wings and a broadening velocity which is estimated in 365$^{+65}_{-69}$ km\,s$^{-1}$ (at $1\sigma$). { The estimate for the broadening velocity was obtained by fitting simultaneously the first order RGS1 and RGS2 spectra from the {\it bapec} model in Xspec -- which describe velocity and thermally-broadened emission from collisionally-ionized gas, taking into account the instrumental broadening. The fit was carried out from photons collected during the whole observations in time intervals free of high particle background, constrained to the oxygen line at the 18--20 \AA\ range. A Gaussian line (added to a bremsstrahlung model to describe the continuum) was used to estimate the flux in the line and the equivalent width for the O\,{\small VIII}\,Ly$\alpha$ line, resulting in $\sim \,1.0 \times 10^{-4} \,{\rm photons \,cm^{-2}\,s^{-1}}$ and $\sim \,30 \, \rm eV$, respectively.}

From Figure~\ref{fig:trailedspec} we note that there is no significant changes in the intensity of the emission lines during times of eclipse with respect to times out of eclipse. Although this result can be due to limitations imposed by the signal-to-noise of the data, it implies that the effect of the lines to the medium resolution spectra cannot be large enough to explain the difference between the spectra seen by the EPIC cameras from times of eclipses and out of them (Fig.~\ref{fig:pn}). This indicates that the drop of up to $\sim$\,40\% of the flux in 0.3-1.4\,keV during eclipses seen by the EPIC cameras is more likely dominated by changes in the continuum. On the other hand, this is also supported by the fact that the strongest emission lines are concentrated in discrete parts in the 14--25\,\AA\ band ($\sim$\,0.5--0.9\,keV) while the changes are seen in the entire spectrum as it is evident from Fig.~\ref{fig:pn}. 

To shed light about the origin of spectral changes observed during the eclipses, we first compute the ratio and the fractional difference between the pn/EPIC spectra S1 and S2: S$_2$/S$_1$ and (S$_2$-S$_1$)/S$_2$, respectively. The result shows that both quantities slightly decrease with the increase in energy. Figure~\ref{fig:pn}-bottom shows the values for S2/S1 (dots), with the respective linear regression represented by the solid line. Next, to evaluate the significance of this trend, we computed fake pn spectra via Xspec using the response matrix (and ancilliary) files constructed from our data. It was assumed a black-body emission with a conservative $k$T of 50\,eV and two conditions to roughly represent S1 and S2 in terms of the observed flux with the only aim of verifying when variations in the black-body component can be detected: (i) changing only the flux of the black-body component (in 15\%) and (ii) changing only the photoelectric absorption (in 5\%). The dashed line in Fig.~\ref{fig:pn}-bottom represents the linear regression from the values of S2/S1 obtained from the fake spectra in the case (i) above. The exposure time assumed to construct the fake spectra was the same as that of the pn data explored in this work. Such exercise proved that the observed spectra S1 and S2 have sufficient signal-to-noise to allow distinction between changes due to variations purely in the flux of the black-body component and changes due purely to variations in the photoelectric absorption at the $3\sigma$ { - for the percentual variations assumed for the properties in the fake spectra}. However, it is not possible to discriminate whether the differences are really due to changes in the photoelectric absorption, in temperature of the black-body (possible in the accretion disk) or from additional contributions (for example, from optically thin plasma that could explain the lines which are observed). This analysis suggests that if Thomson scattering of the photosphere emission from the white dwarf by the ionised matter around the system occurs -- and this should be the case for CAL~87 as it is the case for U\,Sco \citep{Ness12}, it is not the only process at work. This is because the pure Thomson scattering can easily respond for the decreasing in flux observed during eclipses but it does not cause changes in the spectral shape.

\section{Discussion and final remarks}
\label{sct:disc}

The eclipse mapping method is the flagship of this work. It was applied to the X-rays observed from the CBSS CAL~87 by the XMM-{\it Newton} satellite  in order to put constraints on the geometry of the accretion structures of the system. 
In addition, we explore X-ray spectroscopy from a {\it qualitative} approach -- as imposed by the limitation of the available spectral models to well describe the complex X-ray emission of the system -- and X-ray photometry. Also, we discuss observed changes in the orbital period of the system.

It is well known that the eclipse in CAL~87 reduces only partially its flux in X-rays \citep[e.g.,][]{1993PASP..105..863S}. This is shown in Fig. \ref{fig:clfold} for the EPIC/XMM-{\it Newton} data, in which the flux decreases to $\sim\,40\%$. Once the accreting object is completely occulted during the eclipse -- due to the high orbital inclination, the residual X-rays must come from an additional extended component(s):  disk and/or circumbinary corona-like region. The eclipse mapping confirms this expectation showing a considerably large un-eclipsed component.

Furthermore, an extended and symmetric emission is clearly seen in the surface brightness from both the ``soft'' (0.3-0.5\,keV) and ``hard'' (0.5-1.4\,keV) light curves (Fig. \ref{fig:emm}). The map from ``hard'' X-rays differs from that of the ``soft'' X-rays by the presence of two features: a brighter one at the center of the map, slowly fading for outer regions, and another feature in a location which is equivalent to that expected for a bright spot at the intersection between the disk and the ballistic stream trajectory (Fig.~\ref{fig:emm}). In other words, this means that the ``hard'' X-rays become more apparent for larger radius and they are (partially) eclipsed latter on than ``soft'' X-rays. Such results are in line with those obtained directly from the X-ray light curves. While both 0.3-0.5\,keV and 0.5-1.4\,keV light curves seem to initiate the eclipse at the same time (Fig.~\ref{fig:clfold}-top), or at most delayed by $\sim 500\, \rm s$ for the 0.5-1.4\,keV light curve, there is an apparent delay of about 2.5--3.5\,ks till the start of the eclipse in the 0.7-1.4\,keV band (Fig.~\ref{fig:clfold}-middle; see { vertical} lines). On the other hand, hardness ratio computed from the 0.7-1.4\,keV and 0.3-0.5\,keV light curves suggests that the source is slightly hardest during eclipses (Fig.~\ref{fig:clfold}-bottom) { -- the increase has low significance due to the low signal to noise of the data}. Moreover, there is marginal evidence from X-ray spectra extracted during the eclipses and out-of-eclipses (Fig. \ref{fig:pn}) that the hardness of the source slightly increases during the eclipse (see Section~\ref{ssec:xrspct}). As the intensity of the emission lines does not depends in a significant way on the eclipses (Fig. \ref{fig:spec} and \ref{fig:trailedspec}), the changes are more likely dominated by the continuum, suggestive of a cirbumbinary corona  where the emission lines are formed. Such feature could likely be fed by winds, from both the disk and the donor star. In fact, the strong O\,{\small VIII} Ly$\alpha$ line has features which are easily explained by an expanding { (and/or rotating)} medium: velocity broadening and asymmetry in the wings of the line, showing sometimes double-peaked emission { (the reader is referred to \citealt{2009A&ARv..17..309G}, for a more detailed discussion on this topic)}. Together, {\it the pieces of evidence supports that the wind-driven mass transfer scenario is running in CAL~87, with the presence of winds and accretion disk}.

Even though our measured $\dot{P}$ is of the same order of magnitude as those obtained by \citetalias{2007A&A...472L..21O}, our result ($\dot{P} = (+6\pm2) \times 10^{-10}$) is significantly larger ($\dot{P} = (+1.7\pm0.3) \times 10^{-10}$). Since \citetalias{2007A&A...472L..21O} used a different technique (measuring the orbital period of the system with their data alone), it is important to check that our methods are mutually consistent. Repeating our simulation with their data results in $\dot{P} = (+1.5\pm0.7) \times 10^{-10}$, which is consistent with their original measurement. 

Our results then suggest that, in fact, the change in orbital period ($\dot{P}$) in CAL~87 is not constant, contrary to the assumption made here and by \citetalias{2007A&A...472L..21O}. In this case, our result (and that of \citetalias{2007A&A...472L..21O}) can be seen as an average $\dot{P}$.
This behavior of $\dot{P}$ is usually observed on Cataclysmic Variable (CV) like objects (e.g., \citealt{2008A&A...480..481B}), and have been suggested to be caused by circumbinary tertiary companions \citep[and references therein]{2014MNRAS.437..475M,2010MNRAS.407.2362P}. We also point it out that this conclusion should not affect the results of our analysis, since we corrected our timings for this average $\dot{P}$ (see Section~\ref{sct:data}).

Furthermore, it is important to point out that the results of  \citetalias{2007A&A...472L..21O} are based on actual changes in the orbital period of the system and not in eclipse timings. In this case, the variation must be intrinsic to the system, and cannot be associated with the light-travel time effect caused by a tertiary component (e.g., \citealt{2010MNRAS.407.2362P}). In any case, it would be quite difficult to concur with a substellar-mass object been able to sustain a stable orbit in such an extreme environment around CAL~87. 

Even though we do not have enough information to constraint the changes in $\dot{P}$ with time, we can use our result together with those of \citetalias{2007A&A...472L..21O} to draw an overall picture. In this regard, we note that the $\dot{P}$ of \citetalias{2007A&A...472L..21O} was obtained with respect to the timings of \cite{1997MNRAS.287..699A}, covering a time frame of $\sim 10 \rm \, yr$. Our observations, on the other hand, were performed $\sim 7.2 \, {\rm yr}$ after that of \cite{1997MNRAS.287..699A} and $\sim 2.5 \, {\rm yr}$ before those of \citetalias{2007A&A...472L..21O}. We then performed the procedure detailed in Section~\ref{sct:data} using our data as reference and calculate $\dot{P}$ for the data of \citetalias{2007A&A...472L..21O}, which resulted in $\dot{P} = (-2.2\pm0.2) \times 10^{-10}$. Therefore, taking our observations as reference, CAL~87 has a decrease in orbital period. This switch from positive to negative in $\dot{P}$ suggests that CAL~87 should display cyclic orbital period changes. 

The two more likely explanations for cyclic changes are either a circumbinary tertiary component or changes in the angular momentum loss mechanism, modulating the orbital period of the system \citep{1992ApJ...385..621A}. We have already ruled out the circumbinary tertiary component assumption (see earlier discussion). As discussed earlier, CAL~87 is a case where the WDMT scenario is more likely to explain the observed high mass-transfer rate. This implies that the donor sustain a high density and ``strong" stellar wind. At the same time, this component is a fast-rotating low-mass star likely harboring a strong solar-like magnetic field. Therefore, it is possible that magnetic field cycles, as those observed in the the Sun and other stars \citep{1995ApJ...438..269B}, may induce variations in the angular momentum loss rate and, consequently, in the orbital period of the system. It would certainly be helpful for future observations to investigate this effect more thoroughly, which may help to understand the role of cyclic changes on CV and CV-like objects. 

\acknowledgments

The authors thank Jo\~ao E. Steiner for the invaluable comments and the referee for carefully revising the manuscript. This work was supported by the Brazilian agency {\it Instituto Nacional de Ci\^encia e Tecnologia - Astrof\'isica}.

\end{document}